\RequirePackage{fix-cm}
\documentclass[smallextended]{svjour3}       % onecolumn (second format)
\smartqed  % flush right qed marks, e.g. at end of proof
\usepackage{graphicx}
\usepackage[english]{babel}
\usepackage{amsmath}         % Für bessere Mathematikumgebungen
\usepackage{amssymb}         % Für mehr Mathematik Symbole
\usepackage{siunitx}
\usepackage{color} 
\usepackage{listings} 

\begin{document}

\title{
Numerical Simulations of Hyperfine Transitions of Antihydrogen
%Testing Reachable Precision of GS HFS in Antihydrogen within the ASACUSA Setup with Simulations
%Numerical Simulations of Reachable Precision of GS HFS in Antihydrogen within the ASACUSA setup
% 
% Numerical Simulations of 
}


\author{B.~Kolbinger \and
		A.~Capon		\and
		M.~Diermaier \and
		S.~Lehner \and
		C.~Malbrunot \and
		O.~Massiczek \and
        C.~Sauerzopf \and
        M.C.~Simon     \and
        E.~Widmann
}


\institute{B.~Kolbinger \and
		A.~Capon		\and
		M.~Diermaier \and
		S.~Lehner \and
		O.~Massiczek \and
        C.~Sauerzopf \and
         M.C.~Simon    \and
        E.~Widmann \at
              Stefan Meyer Institute for Subatomic Physics, Austrian Academy of Sciences, \\
Boltzmanngasse 2, 1090 Vienna, Austria              \\
              Tel.: +43-(0)1-4277 29710\\
              Fax:  +43-(0)1-4277 9297\\
              \email{bernadette.kolbinger@oeaw.ac.at}             \\              
 C.~Malbrunot \at
             CERN 1211, Geneva 23, Switzerland}

\date{Received: date / Accepted: date}

\maketitle

\begin{abstract}
One of the ASACUSA (Atomic Spectroscopy And Collisions Using Slow Antiprotons) collaboration's goals is the measurement of the ground state hyperfine transition frequency in antihydrogen, the antimatter counterpart of one of the best known systems in physics. This high precision experiment yields a sensitive test of the fundamental symmetry of CPT. Numerical simulations of hyperfine transitions of antihydrogen atoms have been performed providing information on the required antihydrogen events and the achievable precision.
% $\overline{\text{H}}$

\end{abstract}

%%%%%%%%%%%%%%%%%%%%%%%%%%%%%%%%%%%%%%%%%%%%%%%%%%%%%%%%%%%%%%%%%
%%%%%%%%%%%%%%%%%%%%%%%%%%%%%%%%%%%%%%%%%%%%%%%%%%%%%%%%%%%%%%%%%
\section{Introduction}
\label{intro}

\begin{figure}
\centering
  \includegraphics[width=8cm]{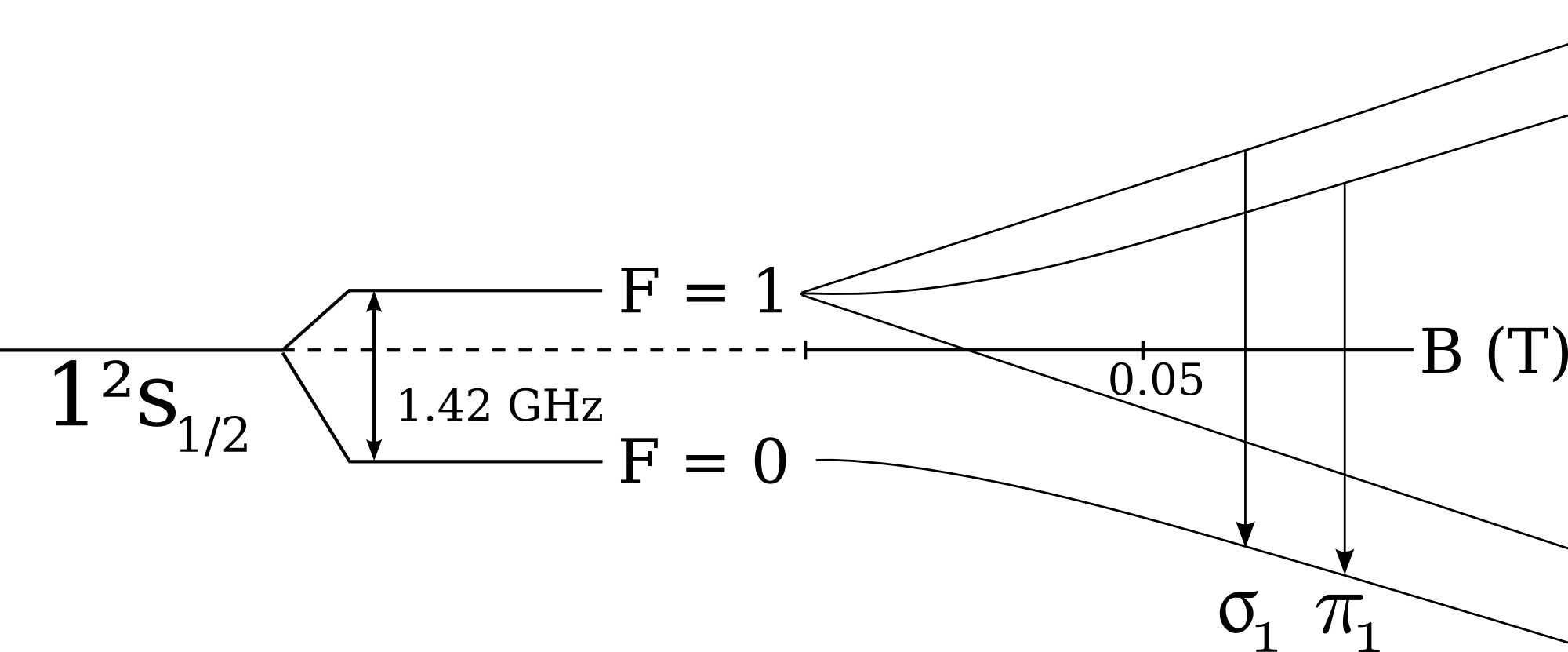}
\vspace*{0.5cm}
\caption{1s groundstate levels of antihydrogen and level splitting in a magnetic field, as well as the two transitions $\pi_1$ $(F=1,M_F=-1) \rightarrow (F=0,M_F=0)$ and $\sigma_1$ $(F=1,M_F=0) \rightarrow (F=0,M_F=0)$.}
\label{fig:1}       % Give a unique label
\end{figure}

Through a precise measurement of the hyperfine splitting of antihydrogen and a comparison with hydrogen, the ASACUSA collaboration aims at testing the CPT symmetry \cite{RefSME}.
In ground state hydrogen the interaction between proton and electron spin leads to a singlet state with quantum number $F=0$ and a triplet state with $F=1$ (see figure \ref{fig:1}). The transition frequency between these two levels is one of the most accurately measured quantities and is therefore well suited to test CPT with very high precision \cite{RefWid,RefWid2}.
%The aim is to achieve a relative precision of $10^{-7}$. In a Rabi-like setup using a beam with a Maxwell-Boltzmann distributed velocity the line width of a resonance spectra can be estimated via \cite{Refwidth}: $f=1.073/T$ with the average time $T$ spent in a radiation field. Together with the double peak structure due to the oscillating B-field configuration of the ASACUSA cavity and an estimated time $T$ of $10^{-4}$s, an precison of $10^{-7}$.
%In a beam setup with a beam velocity of 1000 m/s and a cavity length of 10 cm ($10^{-4}$ s in radiation field), a relative precision of E-7 can be achieved
%%%%%%%%%%%%%%%%%%%%%%%%%%%%%%%%%%%%%%%%%%%%%%%%%%%%%%%%%%%%%%%%%
%%%%%%%%%%%%%%%%%%%%%%%%%%%%%%%%%%%%%%%%%%%%%%%%%%%%%%%%%%%%%%%%%
In the presence of a magnetic field the degenerated $F=1$ level splits up (see figure \ref{fig:1}) and the energies of all four states shift. This is described by the Breit-Rabi formulae  \cite{Refrabi}:
\begin{align}
E_{(1,1)} & = \frac{1}{4} E_0 - \frac{1}{2}(g_J + g_I) \mu_B B \\
E_{(1,0)} & = - \frac{1}{4} E_0 + \frac{1}{2} E_0 \sqrt{1+x^2} \\
E_{(1,-1)} & = \frac{1}{4} E_0 - \frac{1}{2}(g_J + g_I) \mu_B B \\
E_{(0,0)} & = - \frac{1}{4} E_0 - \frac{1}{2} E_0 \sqrt{1+x^2}
\end{align}
where $E_0=h\nu_0$ with the zero magnetic field transition frequency $\nu_0$ and the shifted energies $E_{(1,1)}$ etc. of the four hyperfine levels. $B$ denotes the external magnetic field, $x = B/B_0$ and $B_0 = 2 \pi \nu _0/((g_J - g_I)\mu _B)$ with the electron g-factor $g_J = -2.0023193043718$ \cite{Refvalues} and $g_I = g_p m_e/m_p = 0.003042064412$ \cite{Refvalues} with the proton g-factor $g_p$ (both in units of the Bohr magneton $\mu_B$).
Consequently not only one but several transitions can be observed.
These states can be classified into low- and high-field seekers (LFS and HFS) depending on their behaviour in an inhomogeneous magnetic field. Depending on the alignment of their mangetic moment in a field, atoms with parallel magnentic moment will tend to move toward higher field regions, for states with antiparalell magnetic moment the opposite is the case.  Relevant for the ASACUSA experiment are the $\sigma_1$ and the $\pi_1$ transition:
\begin{align}
\label{breitrabieq}
\sigma _1 : (1,0) \rightarrow (0,0): & \hspace{0.5cm} \nu _{\sigma_1} = \nu _0 \sqrt{1+x^2} \\
\pi _1 : (1,-1) \rightarrow (0,0): & \hspace{0.5cm} \nu_{\pi_1} = \frac{1}{2} \nu _0 - \frac{1}{2}(g_J + g_I) \mu _B B/h + \frac{1}{2} \nu _0 \sqrt{1+x^2}
\end{align}

Numerical simulations of these transitions have been done in order to determine the obtainable precision and will be discussed in the next section.
%The aim is to achieve a relative precision of $10^{-7}$. 
The experimental resolution of the ground state hyperfine transitions is inversely proportional to the interaction time of the atoms with the microwave field in the cavity. The line width of a resonance scan can therefore be estimated by $\delta \nu = 0.799/T$ \cite{Refwidth} where $T$ is the flight time of the atoms through the cavity. The length of the cavity is 10 cm and the average beam velocity is expected to be 1000 m/s which leads to a estimated FWHM of $7 \times 10^{-6}$. With good enough statistics it should be possible to determine the center of the resonance spectrum, which corresponds to the transition frequency, with a relative precision of $10^{-7}$.

%Estimated width of a resonance curve of delta vu ~1/T =... where T is the flight time of the atoms through the cavity.

%the FWHM of a resonance spectra can be estimated via \cite{Refwidth}: $f=1.073 v_m/L$ with the average velocity $v_m$ and the length of the cavity $L$. In a beam setup with an average velocity of 1000 m/s and a cavity length of 10 cm this results in an estimated line width of 10$^{-5}$.  
%The energy of the atom in a B-field shifts according to $\Delta E = - \mu  \vec{B}$ and the magnetic moment will align paralell or antiparallel to the B-field.
\section{Simulations}
\label{exp:sim}

\begin{figure}
\centering
  \scalebox{0.65}{\input{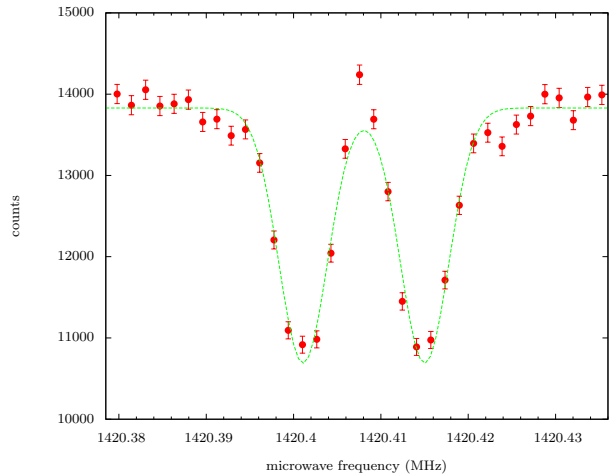}}
\caption{Resonance scan of the $\sigma _1$ transition: count rate on the detector in dependance of microwave frequency. A beam with 3.5$\times 10^6$ detected particles per scan, Maxwell-Boltzmann distributed velocity with a temperature of 50 K, a polarization of 70$\%$ HFS states and 30$\%$ LFS states and a measured fieldmap with $B_{av}^1$ = 8.78$\times 10^{-5}$ T with rms$_1$ = 5.362$\times 10^{-7}$ T.}
%\vspace*{-0.55cm}
\label{fig:2}       % Give a unique label
\end{figure}

The ASACUSA spectrometer line is built up of an antihydrogen source \cite{Refnature}, a spin flip inducing microwave cavity, a spin analyzing sextupole magnet and a detector. The oscillating magnetic field $B_{osc}$ which induces the spin flips is provided by the double stripline resonator of the cavity placed inside a cylindrical vacuum tank \cite{RefSilke}. $B_{osc}$ is very homogeneous in the plane orthogonal to the beam and has a sinusoidal distribution parallel to the beam \cite{Reffield}. As a consequence of this, the resonance spectra of a transition will have a double peak structure leading to zero signal  (i.e. no induced spin flips) when the frequency of the oscillating field is on resonance. This can be seen in the plot of the simulated resonance scan as shown in figure \ref{fig:2}. %\cite{Juhasz2011a}

In the ASACUSA cavity two kinds of transitions are possible -- the $\sigma _1$ and the $\pi _1$ -- depending on the angle between the oscillating magnetic field and the static magnetic field, $B_{stat}$, provided by Helmholtz coils. For the $\sigma _1$ transition, the two B-fields need to be parallel whereas for the $\pi _1$ transition they have to be orthogonal to each other. In the experiment a small static magnetic field is used. Between 0 and 10 Gauss the transition frequency of $\sigma _1$ has a second order dependence on the external field whereas the $\pi_1$ transition varies linearly with the external field and is thus more sensitive to B-field inhomogeneities.

There are now two possible ways of determining the transition frequency at zero static magnetic field $B_{stat}$. Firstly, by measuring the resonance frequency of the $\sigma _1$ transition at different external B-fields $B_{stat} ^i$ and then extrapolating to zero field using the Breit-Rabi formula or secondly, by measuring a resonance scan of both the $\sigma_1$ and $\pi_1$ transition at the same $B_{stat}$ and using the Breit-Rabi formulae in order to extract the hyperfine transition frequency:
%\begin{align}
%E_{(1,1)} & = \frac{1}{4} E_0 - \frac{1}{2}(g_J + g_I) \mu_B B + \Delta E_{(1,1)} \\
%E_{(1,0)} & = - \frac{1}{4} E_0 + \frac{1}{2} E_0 \sqrt{1+x^2} + \Delta E_{(1,0)} \\
%E_{(1,-1)} & = \frac{1}{4} E_0 - \frac{1}{2}(g_J + g_I) \mu_B B - \Delta E_{(1,1)} \\
%E_{(0,0)} & = - \frac{1}{4} E_0 - \frac{1}{2} E_0 \sqrt{1+x^2} - \Delta E_{(1,0)}
%\end{align}
%{\footnotesize
%\begin{align}
%\label{breitrabieq}
%\sigma _1 : (1,0) \rightarrow (0,0): & \hspace{0.5cm} \nu _{\sigma_1} = \nu _0 \sqrt{1+x^2} + \frac{2 \Delta E _{(1,0)}}{h} \\
%\pi _1 : (1,-1) \rightarrow (0,0): & \hspace{0.5cm} \nu_{\pi_1} = \frac{1}{2} \nu _0 - \frac{1}{2}(g_J + g_I) \mu _B B/h + \frac{1}{2} \nu _0 \sqrt{1+x^2} + \frac{\Delta E_{(1,1)} + \Delta E _{(1,0)}}{h} \\
%\pi _2 : (1,0) \rightarrow (1,1): & \hspace{0.5cm} \nu _{\pi_2}= \frac{1}{2} \nu_0(\sqrt{1+x^2}-1) - \frac{1}{2}(g_J + g_I) \mu _B B / h + \frac{\Delta E_{(1,0)} - \Delta E _{(1,1)}}{h}
%\end{align}
%}%
\begin{equation}
\label{eq_sigma_pi}
\nu _0 = \dfrac{g_{+} \sqrt{g_{+}^2 \nu_{\sigma} ^2 - 4 g_{-}^2  \nu_{\pi} ^2 + 4 g_{-}^2  \nu_{\pi}  \nu_{\sigma}} + g_{-}^2(2 \nu_{ \pi} - \nu_{\sigma})}{g_{+}^2 + g_{-}^2}
\end{equation}
where $g_{\pm} = g_I \pm g_J$.
Simulations of the setup are being done using the particle physics toolkit Geant4. It is designed for high energy physics and therefore had to be modified in order to allow simulations of hyperfine transitions in a microwave field. The evolution of the spin state of the atoms for a certain time in the radiation field of the cavity is determinated by solving the optical Bloch equations using a Runge Kutta algorithm with adaptive stepsize control.
%%%%%%%%%%%%%%%%%%%%%%%%%%%%%%%%%%%%%%%%%%%%%%%%%%%%%%%%%%%%%%%%%
%%%%%%%%%%%%%%%%%%%%%%%%%%%%%%%%%%%%%%%%%%%%%%%%%%%%%%%%%%%%%%%%%
One of the simulated resonance scans can be seen in figure \ref{fig:2}.
For reasons of simplicity two gaussians were used as a fit function to determine the center position which corresponds to the transition frequency. A comparison of this simple fit function with a more sophisticated fit algorithm showed no significant difference in precision and value for the center frequency, hence the simpler fit is sufficient for the goal of the present study.

\begin{figure}
\centering
  \scalebox{0.65}{\input{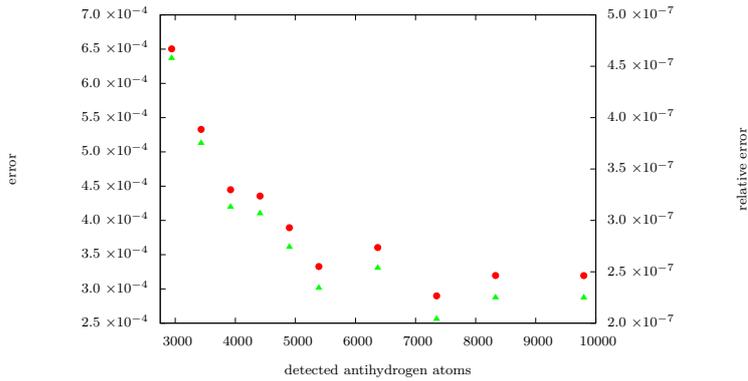}}
\caption{Extrapolation to 0 field using $\sigma_1$ simulations at 3 different fields and 35 points per scan: Absolute (points) and relative errors (triangles) depending on particle number per scan. Two gaussians were used as a fit function.}
\label{fig:3}       % Give a unique label
\end{figure}

%dummy detector, shot directly at cavity,  transfreq rein tun?
%\begin{table}
%\centering
% \begin{tabular}{|c|c|}
% \hline
%    Mean (T) & RMS (T) \\
%   \hline
%    8.78 $\times 10^{-5}$ &  5.362 $\times 10^{-7}$ \\ %5.32120491252
%    2.194 $\times 10^{-4}$ & 1.414 $\times 10^{-6}$  \\ %0.000197694735309
%    4.389 $\times 10^{-4}$ & 2.911 $\times 10^{-6}$  \\ %0.000197694735309
%   \hline
% \end{tabular}
% \caption{Mean magnetic field und RMS of the measured fieldmaps used in the simulations.}
% \label{table_fs}
% \end{table}
%  
The two possible ways of determining the zero field transition frequency have been compared. When only using the $\sigma_1$ transition, scans have been simulated at three different static magnetic fields $B_{stat}^i$ using measured fieldmaps inside the cavity, in order to make the simulations as realistic as possible. The mean field strength and RMS of the measured fieldmaps used are: $B_{av}^1$ = 8.78$\times 10^{-5}$ T with rms$_1$ = 5.362$\times 10^{-7}$ T,
 $B_{av}^2$ =  2.194$\times 10^{-4}$ T with rms$_2$ = 1.414$\times 10^{-6}$ T and
 $B_{av}^3$ = 4.389$\times 10^{-4}$ T with rms$_3$ = 2.911$\times 10^{-6}$ T.
A beam with a Maxwell-Boltzmann distributed velocity of 50 K and a polarization of 70$\%$ LFS and 30$\%$ HFS has been used, since this is the expected polarization of ASACUSA's antihydrogen source \cite{Refcusp}.
\\
For the second method, the same beam properties were used. The angle between $B_{stat}$ and $B_{osc}$ was changed to 45$^{\circ}$ in order to have a parallel and a orthogonal component to $B_{osc}$ and being able to drive both resonances in the same configuration.
\\
With very high statistics, (3.5$\times 10^6$ atoms per scan detected) as in figure \ref{fig:2}, the method using only the $\sigma _1$ transition showed a larger error by a factor of $\approx$1.12. When using lower statistics the method using only the $\sigma_1$ transition becomes more favorable due to the sensitivity on inhomogeneities of the $\pi_1$ transition. Simulations with 2.1$\times 10^5$ particles per scan reaching the detector, which is the particle number where the sensitive $\pi_1$ transition becomes visible, have been done. Comparing the relative error of the two methods shows that the second one yields a higher relative error by a factor of $\approx$1.4 for the homogeneity of the previously listed measured field maps. This can be improved by decreasing the inhomogeniety of B$_{stat}$ which is already planned.
%6000 which is the number of particles per scan point where the sensitive π1 transition becomes clear
\\
In figure \ref{fig:3} the achieved precision of the zero magnetic field HFS is plotted as a function of the number of detected particles. Here, the method of extrapolating three $\sigma_1$ transitions at different $B_{stat}^i$ was used.
\\
Simulations show that the lowest possible particle number at which the double peak structure is still visible is about 2800 detected atoms per scan when using 35 points yields a relative precision of $\approx$ 4.5$\times 10^{-7}$. The relative error corresponding to 3.5$\times 10^{6}$ detected atoms per scan improves this by an order of magnitude.
%%%%%%%%%%%%%%%%%%%%%%%%%%%%%%%%%%%%%%%%%%%%%%%%%%%%%%%%%%%%%%%%%
%%%%%%%%%%%%%%%%%%%%%%%%%%%%%%%%%%%%%%%%%%%%%%%%%%%%%%%%%%%%%%%%%
%Comparing the width of one of the gaussians and the precision shows that the precision is lower a factor of about 10.
%width : 0.006 MHz  relat: 4.2 10 - 6
%precis : 0.0005 MHz  relat: 3.5 10 -7 
%It seems as if the achievable relative resolution levels off at a few $10^{-7}$. The results shown where done with one resonance scan per particle number therefore simulating more scans with the same statistics could clarify this behaviour.
%%%%% comment on level off... 
%%%%%%%%%%%%%%%%%%%%%%%%%%%%%%%%%%%%%%%%%%%%%%%%%%%%%%%%%%%%%%%%%
%%%%%%%%%%%%%%%%%%%%%%%%%%%%%%%%%%%%%%%%%%%%%%%%%%%%%%%%%%%%%%%%%
\section{Summary}
\label{sec:3}
Results of numerical simulations of hyperfine transitions of antihydrogen within the ASACUSA cavity have been presented. 
The zero static magnetic field transition frequency has been determined by two different methods using the $\sigma_1$ and the $\pi_1$ transition frequencies and the resulting precisions have been compared. Simulations show that using both transitions only leads to a higher precision when using high statistics due to the sensitivity to magnetic field inhomogeneities of the $\pi_1$ transition.
The impact of particle numbers on the determined hyperfine transition frequency and its relative error has been discussed. Simulations show that the minimum particle number needed, in order to see the characteristic double peak structure, is 2800 atoms per scan and the achievable relative error $\approx$ 4.5 $\times 10^{-7}$. 

\section{Acknowledgements}
This work is supported by the European Research Council grant no. 291242-HBAR-HFS and the Austrian Ministry for Science and Research.

\end{document}